\documentclass[a4paper,12pt]{article}  
\usepackage[dvips]{graphicx}
\usepackage{amssymb}
\usepackage{cite}
\newcommand{\goes}{\rightarrow} 
 
\newcommand{\GeV}{\; \mathrm{GeV}} 
\newcommand{\TeV}{\; \mathrm{TeV}} 
\newcommand{\lapproxeq}{\lower .7ex\hbox{$\;\stackrel{\textstyle  
<}{\sim}\;$}} 
\newcommand{\gapproxeq}{\lower .7ex\hbox{$\;\stackrel{\textstyle  
>}{\sim}\;$}} 
\newcommand{\stackdown}[2]{\lower 1.4ex\hbox{$\;\stackrel{\textstyle{#1}}  
{\scriptstyle{#2}}\;$}}
\newcommand{\beq}{\begin{equation}} 
\newcommand{\eeq}{\end{equation}} 
\newcommand{\bea}{\begin{eqnarray}} 
\newcommand{\eea}{\end{eqnarray}}
\newcommand{\lsp}{\tilde{\chi}}
\newcommand{\mlsp}{m_{\lsp}}

\newcommand{\relic}{\Omega_{\lsp}\,h_0^2} 
 
\newcommand{\etal}{\textit{et. al.}}
\newcommand{\almuon}{\alpha_{\mu}^{\mathrm{SUSY}}}       
                     
\newcommand{\bsga}{b \goes s \, \gamma}

\makeatletter 
\def\slash{\@ifnextchar[{\fmsl@sh}{\fmsl@sh[0mu]}} 
\def\fmsl@sh[#1]#2{%
  \mathchoice 
    {\@fmsl@sh\displaystyle{#1}{#2}}%
    {\@fmsl@sh\textstyle{#1}{#2}}%
    {\@fmsl@sh\scriptstyle{#1}{#2}}%
    {\@fmsl@sh\scriptscriptstyle{#1}{#2}}} 
\def\@fmsl@sh#1#2#3{\m@th\ooalign{$\hfil#1\mkern#2/\hfil$\crcr$#1#3$}} 
\makeatother 

\begin{document} 
\begin{titlepage} 
 
\begin{flushright} 
\parbox{4.6cm}{hep-ph/XXXXXXX\\
               MIFP-03-06\\
               ACT-03-02 }
\end{flushright} 
\vspace*{5mm} 
\begin{center} 
{\large{\textbf {WMAPing out Supersymmetric Dark Matter and Phenomenology  
}}}\\
\vspace{14mm} 
{\bf A.~B.\ Lahanas} $^{1}$ \,,and, 
{\bf D.~V.~Nanopoulos} $^{2}$   

\vspace*{6mm} 
 $^{1}$ {\it University of Athens, Physics Department,  
Nuclear and Particle Physics Section,\\  
GR--15771  Athens, Greece}

\vspace*{6mm} 
$^{2}$ {\it George P. and Cynthia W. Mitchell 
                Institute of Fundamental Physics, \\   
     Texas A \& M University, College Station,  
     TX~77843-4242, USA \\[1mm]
     Astroparticle Physics Group, Houston 
     Advanced Research Center (HARC),\\ 
     The Mitchell Campus, The Woodlands, TX 77381, USA  \\[1mm] 
     Chair of Theoretical Physics,  
     Academy of Athens,  
     Division of Natural Sciences, 28~Panepistimiou Avenue,  
     Athens 10679, Greece }

\end{center} 
\vspace*{15mm} 
\begin{abstract}
The recent WMAP data provide a rather restricted range of the 
Cold Dark Matter (CDM) density $\; \Omega_{CDM}\; h^2 \;$ of 
unprecedented accuracy. We combine these new data along with data 
from BNL  E821  experiment measuring $\;{(g_{\mu}-2)}$ , $\;$the \\ 
{$b\goes s \;\gamma$} branching ratio and the light 
Higgs boson mass bound from LEP, to update our analysis of the 
allowed boundaries in the parameter space of the Constrained Minimal 
Supersymmetric Standard Model ( CMSSM ). The prospects of measuring 
Supersymmetry at LHC look like a very safe bet, and the potential 
of discovering SUSY particles at a $\;\sqrt{s} \;=\;1.1\; \mathrm{TeV}\;$ linear 
collider is enhanced considerably. The implications for Dark Matter 
direct searches are also discussed.
\end{abstract} 

\end{titlepage} 

\newpage 
\baselineskip=18pt 

\section{Introduction}
As promised, the recent WMAP satellite data \cite{wmap} provide estimates of the 
relevant parameters characterizing the Standard Cosmological Model with unprecedented 
accuracy. The WMAP satellite becomes the LEP of the Sky ! The current energy density 
of the Universe is found to be  about $73 \%$ Dark Energy and $27 \%$ Matter. Most of 
the matter density is in the form of Cold Dark Matter (CDM) as only a very tiny 
component may be due to hot neutrino dark matter and warm dark matter is ruled out 
due to the detected early re-ionization of the Universe at a redshift 
$\; z \approx 0.20 \;$. According to WMAP \cite{wmap} the total Dark Matter density is 
$\;\Omega_m h^2 = 0.135^{+ 0.008}_{-0.009} \;$  while the baryon density is 
$\;\Omega_b h^2 = 0.0224 \pm 0.0009\;$. We thus deduce the value for the CDM density,  
allowing $\;2 \sigma \;$, to be  
\bea
\Omega_{CDM} h^2 = 0.1126^{+ 0.0161}_{-0.0181} \,  \label{newvalue}
\eea
not in disagreement with similar recent observations \cite{cmb} but dramatically more 
precise. These new data should sound like music to the ears of the Supersymmetry 
practitioners that have envisioned such a state of affairs for some time now 
\cite{Hagelin}. Let us see why ? 
One of the major and rather
unexpected predictions of Supersymmetry (SUSY), broken
at low energies $M_{SUSY} \thickapprox \mathcal{O}(1 \TeV) $,
while $R$-parity is conserved, is the existence of a stable, neutral
particle, the lightest neutralino  ($\lsp$), 
referred to as the LSP \cite{Hagelin}.
Such particle is an ideal candidate 
for the Cold Dark Matter in the Universe \cite{Hagelin}, 
and much in need now \cite{wmap}. 
Such a prediction fits well with the fact that 
SUSY is not only  indispensable in 
constructing consistent string
theories, but
it also seems unavoidable at low energies ($\sim 1 \TeV$)
if the gauge hierarchy problem is to
be resolved.
Such a resolution provides a measure of the SUSY
breaking scale $M_{SUSY} \thickapprox \mathcal{O}(1 \TeV) $. 
There is  indirect evidence for such a
low-energy supersymmetry breaking scale, from the unification
of the gauge couplings \cite{Kelley} and from the apparent lightness
of the Higgs boson as determined from precise electroweak measurements,
mainly at LEP \cite{EW}. 
In addition, the BNL E821 experiment \cite{E821} delivered
a more precise measurement for the anomalous magnetic moment of the muon
\bea
\alpha_\mu^{\mathrm exp}=11659203(8)\times 10^{-10} \, , 
\eea 
where $\alpha_\mu = (g_\mu-2)/2$, 
and a detailed calculation 
of the hadronic vacuum polarization contribution
to this moment appeared \cite{thomas}
while a  similar 
calculation\cite{davier} based on low-energy $e^+e^-$ 
data drew similar conclusions.
This calculation, especially using inclusive data,
favours  smaller values of the hadronic vacuum polarization.
As a result, the discrepancy between the the Standard Model (SM)
theoretical prediction  and the experimental value for  the  
anomalous magnetic moment of the muon, becomes significant \cite{thomas}
\bea
\delta \alpha_{\mu} = (361 \pm 106) \times 10^{-11} \, ,
\eea
which corresponds to a $3.3 \, \sigma$ deviation.
Combining this with the new WMAP value (\ref{newvalue}) for supesymmetric
dark matter density \cite{wmap}
one restricts  considerably the parameter
space of  the  Constrained Minimal 
Supersymmetric Standard Model (CMSSM). 
In our analysis we take into account some
other important constraints: the branching ratio for
the $\bsga$ transition and the light Higgs mass bound
$m_{h} \geq 113.5 \GeV$  provided by LEP \cite{LEP,cleo}.
Concerning the $\bsga$ branching ratio the $2\, \sigma$
bound $1.8 \times 10^{-4} < BR(\bsga) < 4.5 \times 10^{-4}$ is
used \cite{cleo}.
\section{Neutralino relic density}

We have repeatedly pointed out in the past \cite{LNS,LNSd,LNSd2} 
the importance of the smallness of the 
Dark Matter (DM) relic density in constraining supersymmetric predictions and at the 
same time we have paid special attention to the large $\; \tan \beta \; $ regime.
In this region the neutralino ($\lsp$) pair annihilation 
through $s$-channel
pseudo-scalar Higgs boson ($A$) 
exchange, leads to an enhanced annihilation cross sections
reducing significantly the relic 
density \cite{Drees}. 
The importance of this mechanism, in conjunction with the 
cosmological data which favour small values of the DM
relic density,
has been stressed in \cite{LNS,LNSd}.
As mentioned above for large $\tan \beta$ 
the $\lsp \, \lsp \stackrel{A}{\goes} b \, \bar{b}$ or $\tau \, \bar{\tau}$ 
channel becomes the dominant annihilation mechanism. 
In fact by increasing $\tan \beta$ the mass $m_A$ decreases, while the
neutralino mass remains almost constant, if the other parameters are kept
fixed. Thus  $m_A$ is expected eventually to enter into the regime in which
it is close to the pole value $m_A\,=\, 2 m_{\lsp}$, and the
pseudo-scalar  Higgs exchange dominates.
In previous analyses regarding 
DM direct searches \cite{LNSd}, we had stressed 
that  the contribution of the $CP$-even Higgs bosons exchange
to the LSP-nucleon scattering cross sections increases with $\tan \beta$.
Therefore in the large $\tan \beta$ 
region one obtains the highest possible rates
for the direct DM searches and the smallest LSP relic densities. 
Similar results are presented in Ref.~\cite{Kim}. 
As mentioned before in view of the recent WMAP cosmological data, which 
point towards even smaller and more accurate  values of the CDM abundance, 
the updating of the supersymmetric restrictions imposed is highly 
demanding. In conjuction with the  ($g_\mu -2$) E821 data, which has been  
the subject of intense phenomenological 
study the last couple of years \cite{Ellis,ENO,LS,LNSd2,g-2,Leszek,Kneur},
it may considerably limit the bounds of sparticle masses which is of paramount 
importance for the next run experiments at LHC or other accelerators \cite{ellis2}.
{\footnote{Similar conclusions have been also reached in Ref. \cite{ellis2}.}  

For the correct calculation of the neutralino relic density  
in the large $\tan\beta$ region, 
 an unambiguous and
reliable determination of the $A$-mass is required.
The  details of the 
procedure in calculating the spectrum of the CMSSM can be
found elsewhere \cite{LS,LNSd2}. Here 
we shall only briefly 
refer to some subtleties which turn out to be
 essential for a correct determination of $m_A$.
In the CMSSM, 
$m_A$ is not a free parameter but 
is determined once the other parameters  are given.
$m_A$  depends sensitively on the Higgs
mixing parameter, $m_3^2$, which is determined from minimizing 
the  one-loop corrected effective potential.
For large $\tan \beta$ the derivatives of the effective potential
with respect the Higgs fields, which enter into the minimization conditions, 
are plagued by terms which are large and hence potentially dangerous, making
the perturbative treatment untrustworthy.
In order to minimize the large $\tan \beta$
corrections we had better calculate the effective potential using as
reference scale the average stop scale
$Q_{\tilde t}\simeq\sqrt{m_{{\tilde t}_1} m_{{\tilde t}_2} }$ \cite{scale}. 
At this scale these terms are small and hence perturbatively valid.
Also for the calculation of the pseudo-scalar Higgs boson mass
 all the one-loop corrections must be taken into account. 
In particular, the inclusion of  those of the neutralinos and charginos  
yields  a result for $m_A$ that is scale independent and 
approximates the pole mass to better than $2 \%$  \cite{KLNS}.
A more significant correction, which 
drastically affects the pseudo-scalar mass 
arises from the gluino--sbottom and chargino--stop corrections to the bottom
quark Yukawa coupling  $h_b$ \cite{mbcor,wagner,BMPZ,arno}.
The proper resummation of these corrections
is important for a correct determination of $h_b$ \cite{eberl,car2},
and accordingly of the $m_A$.
Seeking a precise determination  of the Higgs boson mass
the dominant two-loop corrections to this have been included \cite{zwirner}.
Concerning the calculation of the $\bsga$ branching ratio,
the important contributions beyond the leading order, 
especially for large $\tan \beta$ have been taken into account \cite{gamb}. 

In calculating the $\lsp$
relic abundance, we solve the Boltzmann equation  
numerically using the machinery
outlined in Ref.~\cite{LNS}. In this calculation the coannihilation effects, 
in regions where $\tilde{\tau}_R$ approaches in mass the LSP, which is a high
purity Bino, are properly taken into account.

\begin{figure}[t!] 
\begin{center}
\includegraphics[scale=1.25]{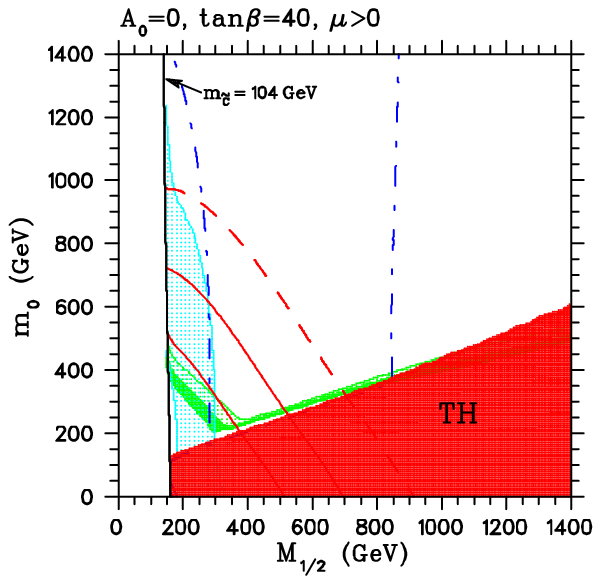}
\hspace*{.1cm}
\includegraphics[scale=1.25]{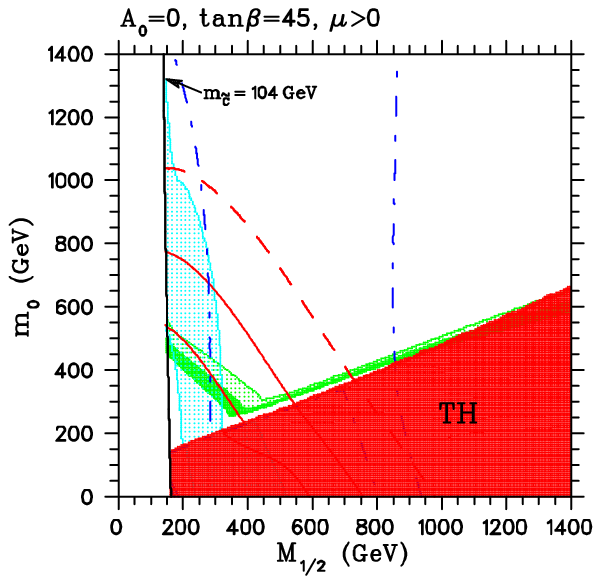}
\end{center}

\caption[]{
Cosmologically allowed regions of the relic density for 
 of $\tan \beta=40$ and $45$ in the $(M_{1/2},m_0)$ plane.  
The mass of the top is taken $175\GeV$. In the dark
green shaded area $0.094<\relic<0.129$. In the light green shaded
area $0.129<\relic<0.180\;$. The solid red lines mark the region within which
the supersymmetric contribution to the anomalous magnetic moment of the
muon is
$\alpha^{SUSY}_{\mu} = (361 \pm 106) \times 10^{-11}$.
The dashed red line
is the boundary of the region for which the lower bound is moved to
its $2 \sigma$ limit. 
The dashed-dotted blue lines are
the boundaries of the region $113.5 \GeV \leq m_{Higgs} \leq 117.0 \GeV$.
The cyan shaded region is excluded due to
$\bsga$ constraint. 
}

\label{fig1}  
\end{figure}

Using the new WMAP value for CDM density (\ref{newvalue}) and the  
bound for $( g_\mu-2 )$ as described in the introduction, the 
parameter space is constrained significantly , as
depicted in the figures. 
In the panels of figure~\ref{fig1} we display our results by drawing 
the cosmologically $\; 2\sigma\;$ 
allowed region $0.094<\relic<0.129$ (dark green), in the $m_0, M_{1/2}$ plane,  
for values of $\tan \beta$ equal to  $40$ and $45$ respectively.
For comparison also drawn, in light green, is the region $0.129<\relic<0.180$.
Note that the value $\;0.180\;$ was our previous upper bound 
\cite{LNS,LNSd,LNSd2}.  
In the figures shown  we used for the top, tau and bottom masses
the default values  
$M_t = 175 \GeV, M_{\tau} = 1.777\GeV$ and $m_b(m_b) = 4.25\GeV$. 
We have fixed $A_0=0$, since our results are not sensitive to
the value of the common trilinear coupling.
The solid red lines mark the region within which
the supersymmetric contribution to the anomalous magnetic moment of the
muon falls within the E821 range 
$ \alpha^{SUSY}_{\mu} = ( 361 \pm 106 ) \times 10^{-11}$.
The dashed red line
marks the boundary of the region when the more relaxed $2\, \sigma$ 
value on the lower bound of the E821 range is used. 
Along the blue dashed-dotted contour lines the light $CP$-even Higgs mass takes
values $113.5\GeV$ (left) and $117.0\GeV$ (right) respectively.
The line on the left
marks therefore the recent LEP bound on the Higgs mass \cite{LEP}.
Also shown is the chargino mass bound $104\GeV$.
The shaded area (in red)
at the bottom of each figure, labelled by TH, is theoretically forbidden  
since the light stau is lighter than the lightest of the neutralinos.
The cyan shaded region is excluded by the $\bsga$ constraint.

\begin{figure}[t] 
\begin{center}
\includegraphics[scale=1.25]{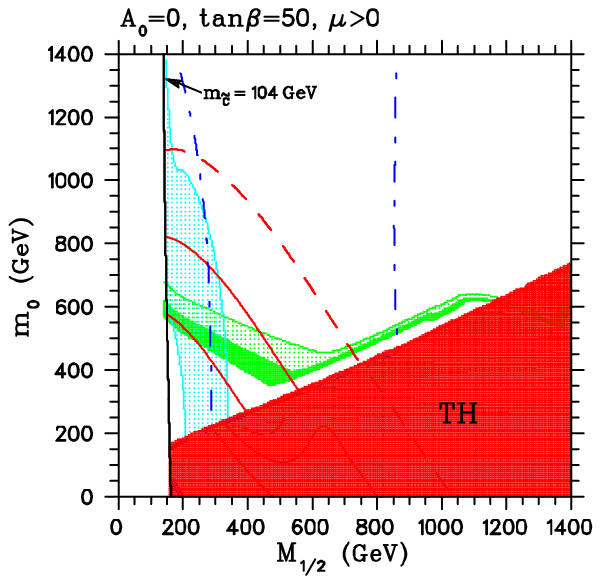}
\hspace*{.1cm}
\includegraphics[scale=1.25]{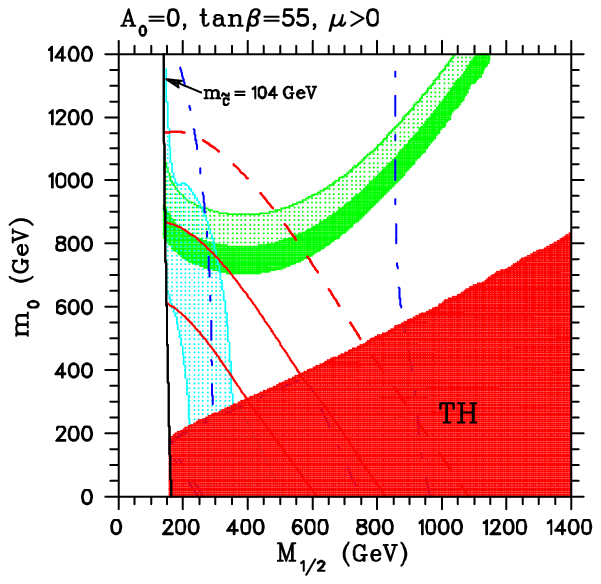}
\end{center}

\caption[]{The same as in Fig.~\ref{fig1} for  $\tan \beta=50$ and $55$. 
}

\label{fig2}  
\end{figure}

For large values of $\tan\beta$,
see the right panel of figure~\ref{fig2},
a region opens up within which the relic density
is cosmologically allowed. This is due to the pair annihilation of the
neutralinos through the pseudo-scalar Higgs exchange in the $s$-channel.
As explained before, for such high $\tan \beta$ the ratio $m_A / 2 m_{\lsp}$
approaches unity and the pseudo-scalar exchange dominates yielding
large cross sections and hence small neutralino relic densities. It is for this 
reason that we give special emphasis to this particular mechanism which opens 
up for large $\; \tan \beta \;$ and delineates cosmologically allowed 
domains of small relic densities and large elastic neutralino - nucleon cross. 
In this case the lower bound put by the $( g_\mu -2 )$ data 
cuts the cosmologically allowed
region which would otherwise allow for very large values of $m_0, M_{1/2}$.

\begin{table}[t]
\begin{center}
\begin{tabular}{|c|c|c|c|c|c|} \hline \hline
 $\tan\beta$ & $\lsp^0$ & $\tilde{\chi}^+$ & $\tilde{\tau}$ & $\tilde{t}$ & $h$ 
                                                                    \\ \hline
  10  &  155  & 280  & 170 & 580  &  116 \\
  15  &  168  & 300  & 185 & 640  &  116 \\
  20  &  220  & 400  & 236 & 812  &  118 \\
  30  &  260  & 470  & 280 & 990  &  118 \\
  40  &  290  & 520  & 310 & 1080 &  119 \\
  50  &  305  & 553  & 355 & 1120 &  119 \\
  55  &  250  & 450  & 585 & 970  &  117 \\ 
 \hline \hline
\end{tabular}
\end{center}

\vspace{.4cm}
\caption{Upper bounds, in GeV, 
on the masses of the lightest of the neutralinos,
charginos, staus, stops and  Higgs bosons for various values of
$\tan\beta$ if the new WMAP value \cite{wmap} for $\Omega_{CDM} h^2$ 
and the $2 \sigma$  E821 bound, 
$149 \times 10^{-11}<\alpha_{\mu}^{SUSY}<573 \times 10^{-11}$,  
is imposed
}
\label{table1}
\end{table}

For the $\tan \beta = 55$ case, 
close to the highest possible value, and considering the 
$2 \, \sigma$ lower bound on the muon's anomalous magnetic moment
$\alpha_{\mu}^{SUSY} \geq 149 \times 10^{-11}$ and values of
$\relic$ in the range $\;0.1126^{+ 0.0161}_{-0.0181}  \;$, 
we find that the allowed points are within a narrow stripe.  
The point with the highest value for $m_0$ is ( in GeV ) 
at $(m_0, M_{1/2}) \;=\; ( 850 , 550 ) \;$ and that with the highest 
$\; M_{1/2}\;$ at $(m_0, M_{1/2}) \;=\; ( 750 , 600 \;) \;$. The latter 
marks the lower end of the line segment of the boundary 
$\; 149 < 10^{-11}\;\alpha_{\mu}^{SUSY}\;$ which amputates the 
cosmologically allowed stripe.  
It should be noted that within $1 \sigma$ 
of the E821 data only a few points survive which lie in a small region 
centered at $(m_0, M_{1/2}) \;=\; ( 725 , 300 ) \;$.
The bounds on $m_0, M_{1/2}$ displayed in figure~\ref{fig2} refer to the 
$\; A_0=0 \;$ case.
Allowing for $A_0 \neq 0$ values, the upper bounds put on $m_0, M_{1/2}$
increase a little and so do the corresponding bounds on sparticle
masses.

For the LSP, the lightest of the charginos, stops, staus and Higgses the upper 
bounds on their masses are displayed in Table \ref{table1} for various values 
of the parameter $\; \tan \beta \;$, if the new WMAP determination \cite{wmap} 
of the Dark Matter (\ref{newvalue}) and the $2 \sigma$ bound 
$\; 149 < 10^{-11}\;\alpha_{\mu}^{SUSY}<573\;$ of E821 is respected.  
We have also taken into account the limits arising from Higgs boson searches 
as well as from $\bsga$ experimental constraints. 
In extracting these values we used a random sample  
of $40,000$ points in the region $|A_0| < 1 \; \mathrm{TeV}$, $\tan \beta < 55 $, 
$m_0,M_{1/2} < 1.5 \; \mathrm{TeV}$ and $\; \mu > 0 \;$. 
The lightest of the charginos has a mass 
whose upper bound is $\; \approx 550 \; \mathrm{GeV} \;$, 
and this is smaller than the upper bounds put on the masses of the 
lightest of the other charged sparticles, namely the stau and stop, as is 
evident from Table \ref{table1}. 
Hence the prospects of discovering CMSSM 
at a $e^{+} e^{-}$ collider with center of mass energy $\sqrt s = 800 \GeV$,
are  {\em not} guaranteed. Thus a center of mass energy of at least 
$\sqrt s \approx 1.1 \; \mathrm{TeV}\;$
is required to discover SUSY through chargino pair production.  
Note that in the allowed regions 
the next to the lightest neutralino, ${\tilde{\chi}^{\prime}}$, has a mass
very close to the lightest of the charginos and hence the process
$e^{+} e^{-} \goes {\tilde{\chi}} {\tilde{\chi}^{\prime}}$, with 
${\tilde{\chi}^{\prime}}$ subsequently decaying to
$ {\tilde{\chi}} + {l^{+}} {l^{-}}$ or 
$ {\tilde{\chi}}+\mathrm{2\,jets}$,  
is kinematically allowed for such large $\tan \beta$, provided
the energy is increased to at least $\sqrt{s} = 860 \GeV$. It should
be noted however that this channel proceeds via the $t$-channel exchange
of a selectron and it is suppressed due to the heaviness of the exchanged
sfermion.
Therefore only if the center of mass energy is increased to 
$\; \sqrt{s} = 1.1 \; \mathrm{TeV} \;$ supersymmetry can be discovered in a $e^{+} e^{-}$ 
collider provided it is based on the Constrained scenario.

   
\begin{figure}[t] 
\begin{center}
\includegraphics[scale=.7]{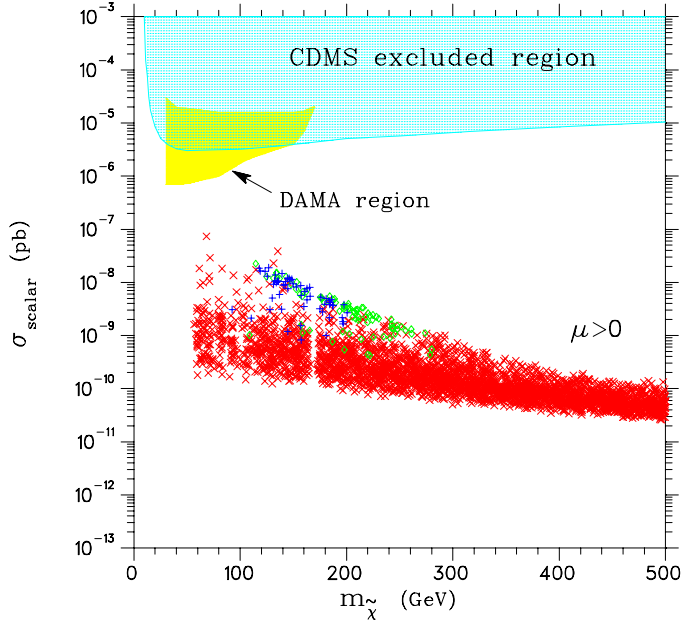}
\end{center}
\caption[]{
Scatter plot of the scalar neutralino-nucleon cross section
versus $\mlsp$, from a random sample of 40,000 points.
On the top of the figure the CDMS excluded region and
the DAMA sensitivity region are illustrated. 
Blue pluses ($+$) are points within the E821 experimental region
$\almuon = ( 361 \pm 106 ) \times 10^{-11}$ 
which are cosmologically acceptable 
$\Omega_{\lsp} h^2 = 0.1126^{+ 0.0161}_{-0.0181}  $.
Green diamonds ($\diamond$)  are points within the $2\sigma$ 
E821 experimental region and also cosmologically acceptable.
Crosses ($\times$) represent the rest of the random sample.
The Higgs boson mass bound $m_h > 113.5 \GeV$ and the $\bsga$
constraint are properly taken into account.}

\label{fig3} 
\end{figure}  

\section{Direct Dark Matter searches}

We turn now to study the impact of the new WMAP data on Dark Matter, 
the $\; ( g_{\mu}-2 )\;$, the $\bsga$ and the Higgs mass bounds, on the direct 
DM searches. The random sample used in this study is the same used in 
the previous section when considering the neutralino relic density. 
In figure~\ref{fig3} we plot the scalar $\lsp$-nucleon   
cross section as function of the LSP mass, $\mlsp$.
On the top of it the shaded region (in cyan colour) is
excluded by the CDMS experiment \cite{cdms}.
The DAMA sensitivity region is
plotted in yellow \cite{dama}. 
Pluses ($+$) (in blue colour) represent points 
which are both compatible with the E821 data 
$\almuon = (361 \pm 106)\times 10^{-11}$
and the WMAP cosmological bounds $\Omega_{CDM} h^2 = 0.1126^{+ 0.0161}_{-0.0181}$.  
Diamonds ($\diamond$) (in green colour) represent points 
which are  compatible with the $2 \, \sigma$ E821 data  
$149 \times 10^{-11}  < \almuon  < 573 \times 10^{-11} $
and the cosmological bounds.
The crosses ($\times$) (in red colour) represent the rest of the
points of our random sample.  Here the Higgs boson
mass, $m_h > 113.5 \GeV$ and the
$\bsga$ bounds have been properly taken into account.
From this figure it is seen that the     
points which are compatible both with the ($g_{\mu}-2 )$ E821
and the cosmological data can yield cross sections
slightly above $10^{-8}$ pb when $\mlsp$
is about $120 \GeV$. The maximum value of $\mlsp$
is around $200 \GeV$ but in this case the scalar cross section drops 
by almost an order of magnitude $10^{-9}$ pb.   
Accepting the $2\, \sigma$ $( g_{\mu}-2 )$ bound the maximum value of the 
scalar cross section is again $\approx 10^{-8}$ pb,  
for $\mlsp \approx 120 \;GeV$, but the 
$\mlsp$ bound is increased to about $280\;GeV$ at the expense of having 
cross sections slightly smaller than $10^{-9}$ pb. 
Considering the $\mu>0$ case,  
it is very important that using all available data, one 
can put a lower bound $\approx 10^{-9}$ pb on the scalar cross section which is 
very encouraging for 
future DM direct detection experiments \cite{Klapdor}. Such a lower bound cannot 
be imposed when $\mu<0$, since the scalar cross section can become very small due 
to accidental cancellations between the sfermion and Higgs exckhange processes. 
However, this case is not favoured by $( g_{\mu}-2 )$ and $\bsga$ data.

\section{Conclusions}
 
We have combined the new WMAP cosmological data \cite{wmap}  
on Dark Matter with recent 
high energy physics experimental information including measurements 
of the anomalous magnetic moment of the muon, from E821 Brookhaven experiment,  
the $\bsga$ branching ratio  and the light Higgs boson mass bound
from LEP and we studied the imposed constraints on the parameter space of the 
CMSSM. We have assessed the potential of discovering SUSY, if it is based on 
CMSSM, at future colliders and DM direct search experiments. The use of the 
new WMAP data in conjuction with the $2 \; \sigma$ $( g_{\mu}-2 )$ bound can guarantee 
that in LHC but also in a $e^{+} e^{-}$ collider with center of mass energy 
$\sqrt{s} \approx 1.1 \; \mathrm{TeV}$ CMSSM can be discovered. The effect of these constraints 
is also significant for the direct DM searches. 
For the $\mu>0$ case we found that the minimum
value of the spin-independent $\lsp$-nucleon
cross section attained is of the order of $10^{-9}$ pb.

\vspace*{1cm}
\noindent 
{\bf Acknowledgements} \\ 
\noindent A.B.L. acknowledges support from HPRN-CT-2000-00148
and HPRN-CT-2000-00149 programmes. He also thanks the University of Athens
Research Committee for partially supporting this work. 
D.V.N. acknowledges support by D.O.E. grant DE-FG03-95-ER-40917.

\end{document}